# Enhancing the Interpretability of SHAP Values Using Large Language Models

Xianlong Zeng[1], Kewen Zhu
[1] xz926813@ohio.edu

## Abstract

Model interpretability is crucial for understanding and trusting the decisions made by complex machine learning models, such as those built with XGBoost. SHAP (SHapley Additive exPlanations) values have become a popular tool for interpreting these models by attributing the output to individual features. However, the technical nature of SHAP explanations often limits their utility to researchers, leaving non-technical end-users struggling to understand the model's behavior. To address this challenge, we explore the use of Large Language Models (LLMs) to translate SHAP value outputs into plain language explanations that are more accessible to non-technical audiences. By applying a pre-trained LLM, we generate explanations that maintain the accuracy of SHAP values while significantly improving their clarity and usability for end users. Our results demonstrate that LLM-enhanced SHAP explanations provide a more intuitive understanding of model predictions, thereby enhancing the overall interpretability of machine learning models. Future work will explore further customization, multimodal explanations, and user feedback mechanisms to refine and expand the approach.

## Introduction

In recent years, machine learning models have grown increasingly complex, often functioning as black-box systems that make critical decisions in various domains, including healthcare, finance, and autonomous systems. While these models can achieve high levels of accuracy, their complexity makes it challenging for stakeholders to understand how specific decisions are made. This lack of transparency can lead to mistrust, especially in high-stakes environments where understanding the reasoning behind a model's predictions is essential for decision-making. As a result, model interpretability has emerged as a vital area of research, aiming to provide insights into the inner workings of these models and ensure that their predictions are understandable and actionable.

SHAP (SHapley Additive exPlanations) values have gained widespread recognition as a powerful tool for model interpretability. By attributing the output of a model to its input features, SHAP values offer a unified measure of feature importance, helping to explain individual predictions. Despite their effectiveness, SHAP values are often presented in a technical manner, making them accessible primarily to data scientists and developers. The granular and mathematical nature of SHAP explanations can be difficult for non-technical users to grasp, limiting their utility in broader applications where end users may lack the technical expertise needed to interpret these outputs.

The growing disconnect between model developers and end users highlights a pressing need for interpretability methods that are both accurate and accessible. To bridge this gap, we propose leveraging Large Language Models (LLMs), which have demonstrated remarkable abilities in generating coherent and contextually relevant natural language explanations. By applying LLMs to SHAP values, we aim to transform these technical explanations into plain language narratives that are easier for non-technical users to understand. This approach not only enhances the interpretability of machine learning models but also democratizes access to model insights, empowering a wider range of users to make informed decisions based on the model's predictions.

In this paper, we explore the application of LLMs to enhance the explainability of SHAP values, providing detailed examples and evaluations of the generated explanations. We demonstrate that LLM-enhanced SHAP explanations retain the accuracy of the original SHAP values while significantly improving their clarity and accessibility. Finally, we discuss the potential implications of this approach for improving trust in machine learning models and outline future directions for further enhancing model interpretability.

As part of our efforts to bridge the gap between technical and non-technical stakeholders in the realm of machine learning interpretability, this paper presents several key contributions:

- **Application of LLMs for SHAP Interpretability**: We introduce the use of Large Language Models (LLMs) to translate SHAP value outputs into plain language explanations, making them more accessible to non-technical users.
- **Demonstration of Enhanced Explanations**: Through examples, we demonstrate that LLM-enhanced SHAP explanations retain the accuracy of the original SHAP values while providing clearer and more understandable insights.
- **Evaluation of Clarity and Accessibility**: We evaluate the effectiveness of LLM-generated explanations in improving the clarity and accessibility of SHAP outputs, highlighting their potential to bridge the gap between technical developers and non-technical end users.
- **Discussion of Future Directions**: We outline potential future enhancements, including customization of LLM outputs, integration with visual aids, and implementation of user feedback mechanisms, to further improve the interpretability of machine learning models.

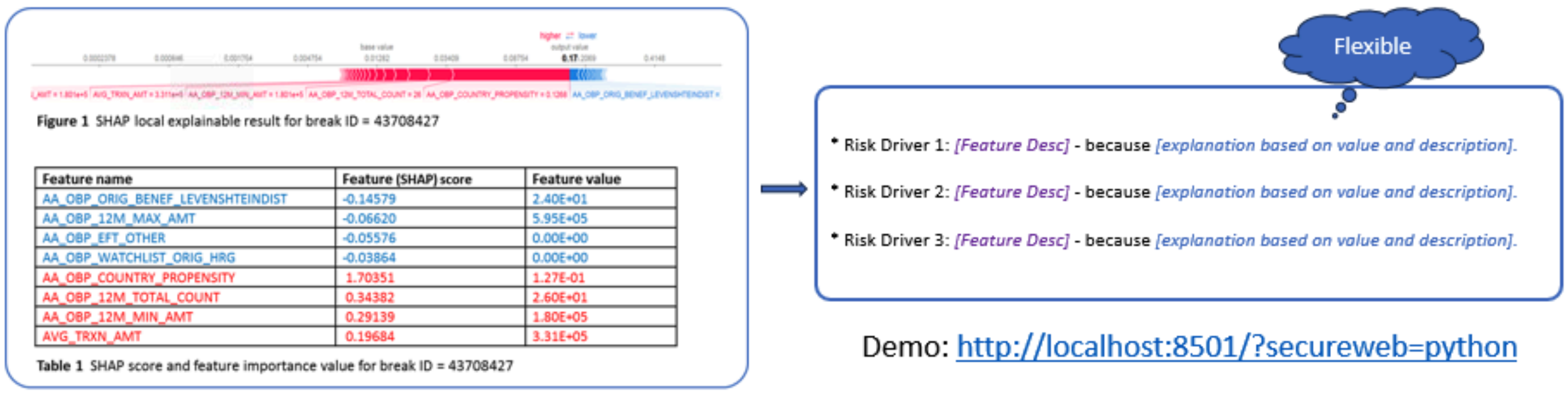

| Feature name | Feature (SHAP) score | Feature value |
|---|---|---|
| AA_OBP_ORIG_BENEF_LEVENSHTEINDIST | -0.14579 | 2.40E+01 |
| AA_OBP_12M_MAX_AMT | -0.06620 | 5.95E+05 |
| AA_OBP_EFT_OTHER | -0.05576 | 0.00E+00 |
| AA_OBP_WATCHLIST_ORIG_HRG | -0.03864 | 0.00E+00 |
| AA_OBP_COUNTRY_PROPENSITY | 1.70351 | 1.27E-01 |
| AA_OBP_12M_TOTAL_COUNT | 0.34382 | 2.60E+01 |
| AA_OBP_12M_MIN_AMT | 0.29139 | 1.80E+05 |
| AVG_TRXN_AMT | 0.19684 | 3.31E+05 |

Table 1 SHAP score and feature importance value for break ID = 43708427

Figure 1. Illustration of Enhancing SHAP Value Interpretability Using Large Language Models

# Methodology

This section outlines the methodology employed to apply Large Language Models (LLMs) for enhancing the interpretability of SHAP (SHapley Additive exPlanations) values. Our approach is designed to translate the often complex and technical outputs of SHAP into plain language explanations that are more accessible to non-technical users.

## Overview of SHAP Values

SHAP values are grounded in cooperative game theory, where the prediction of a machine learning model is viewed as a payout that must be fairly distributed among the input features. The SHAP value for a feature $i$ is calculated as the weighted average of the marginal contributions of $i$ across all possible subsets S of features that do not include $i$:

$$\phi_i = \sum_{S \subseteq N \setminus \{i\}} \frac{|S|!(|N| - |S| - 1)!}{|N|!} [f(S \cup \{i\}) - f(S)] \tag{1}$$

where:

- N is the set of all features,
- S is a subset of features not containing $i$
- f(S) is the model's prediction when only the features in subset S are present.

This equation represents the SHAP value ϕ as the contribution of the feature to the difference in prediction, averaged over all possible combinations of features. While this formula provides a rigorous and fair attribution of the model's output to its inputs, the resulting SHAP values can be difficult for non-experts to interpret, especially when dealing with large models.

## Applying LLMs to SHAP Values

To address the challenge of interpretability, we propose leveraging Large Language Models (LLMs) to generate natural language explanations from SHAP values. The key steps in our methodology are as follows:

- **Input Representation**: SHAP values for a given prediction are extracted and structured as input to the LLM. If we denote the SHAP values for a prediction as $\{\phi_1, \phi_2, \ldots, \phi_n\}$ corresponding to features $\{x_1, x_2, \ldots, x_n\}$, the input to the LLM can be represented as a set of tuples:

$$\text{Input to LLM} = \{(x_1, \phi_1), (x_2, \phi_2), \ldots, (x_n, \phi_n)\} \quad (2)$$

This structured input ensures that the LLM can associate each feature with its corresponding SHAP value, facilitating the generation of meaningful explanations.

- **LLM Selection**: We selected a pre-trained LLM, such as GPT-3 or GPT-4, known for its strong natural language processing capabilities. These models are particularly effective at generating coherent and contextually relevant text based on a given prompt. The choice of LLM was guided by its ability to handle complex technical content and generate clear, concise explanations.
- **Prompt Engineering**: To generate meaningful explanations, we crafted specific prompts that guide the LLM in translating SHAP values into user-friendly language. For example, a prompt might instruct the LLM to "explain the importance of the following features in determining the model's prediction," followed by the list of features and their SHAP values. Prompt engineering plays a critical role in ensuring that the LLM produces explanations that are accurate, relevant, and easy to understand.
- **Generation of Explanations**: Once the SHAP values and prompts are prepared, they are fed into the LLM, which generates natural language explanations. These explanations are intended to be straightforward and provide clear insights into why certain features contributed to the model's prediction. The output might look like a paragraph explaining which features were most influential and why, in terms that a non-technical user can easily comprehend.

## Implementation Details

The implementation of our approach involves deploying a local instance of the Mistral 7B model, a lightweight yet powerful LLM, to generate natural language explanations from SHAP values. The following steps outline the technical setup and process:

- **Environment Setup**: The Mistral 7B model was deployed locally on a machine with sufficient GPU resources to handle its computational requirements. The model was accessed using a Python environment, where it was integrated with other libraries necessary for SHAP value computation and text generation. We utilized the transformers library from Hugging Face to load and interact with the Mistral 7B model. The SHAP values were calculated using the shap library in Python, and the results were passed as inputs to the Mistral model.

- **LLM Integration:** The SHAP values, once calculated for a specific prediction, were structured into a format suitable for input to the Mistral model. The input consisted of feature names and their corresponding SHAP values, organized as tuples. These were then fed into the LLM, accompanied by carefully crafted prompts designed to elicit clear and coherent explanations. The integration was managed through a Python script that automated the entire process, from SHAP value calculation to the generation of natural language explanations.
- **Local Deployment Considerations:** Deploying the Mistral 7B model locally provided several advantages, including reduced latency and greater control over the customization of the model's behavior. Additionally, running the model locally allowed for fine-tuning and optimization specific to the needs of our application. This setup also ensured data privacy, as all processing occurred within the local environment without reliance on external APIs.
- **Post-Processing:** After the Mistral model generated the explanations, a post-processing step was applied to refine the output. This involved basic text cleaning, such as correcting any grammatical errors and removing irrelevant or redundant information. The goal was to ensure that the final explanations were concise, accurate, and easy for non-technical users to understand.
- **Optimization and Tuning:** To further enhance the quality of the generated explanations, we experimented with various prompt structures and input formats. Additionally, hyperparameter tuning was performed on the Mistral model to optimize its performance in generating explanations that align closely with the intended interpretability goals. This tuning process included adjusting the temperature and max tokens settings to balance the verbosity and coherence of the output.

## Example Workflow

To illustrate the methodology, we use the Titanic dataset, a well-known dataset in the machine learning community, to showcase how our approach works:

- **Model Prediction**: An XGBoost model is trained on the Titanic dataset to predict whether a passenger survived or not based on features such as age, gender, passenger class, and fare paid. For a specific passenger, the model predicts the likelihood of survival.
- **SHAP Value Calculation**: SHAP values are computed for the prediction, indicating how much each feature contributed to the model's decision. For example, the SHAP values might show that being a female passenger and traveling in first class had a positive impact on the survival prediction, while older age had a negative impact.
- **LLM Explanation Using Mistral 7B**: The SHAP values, along with the corresponding feature names, are fed into the locally deployed Mistral 7B model. A crafted prompt guides the model to generate a plain language explanation. For instance, the model might output: "The model predicts a high likelihood of survival for this passenger primarily because she is a female traveling in first class, which historically had a higher survival rate. However, the passenger's older age slightly reduces this likelihood."

- **User Output**: The final explanation is presented to the user, detailing the most important features and how they influenced the prediction in an easily understandable format. This output allows non-technical users to comprehend why the model made its prediction for this particular passenger, based on key factors like gender, class, and age.

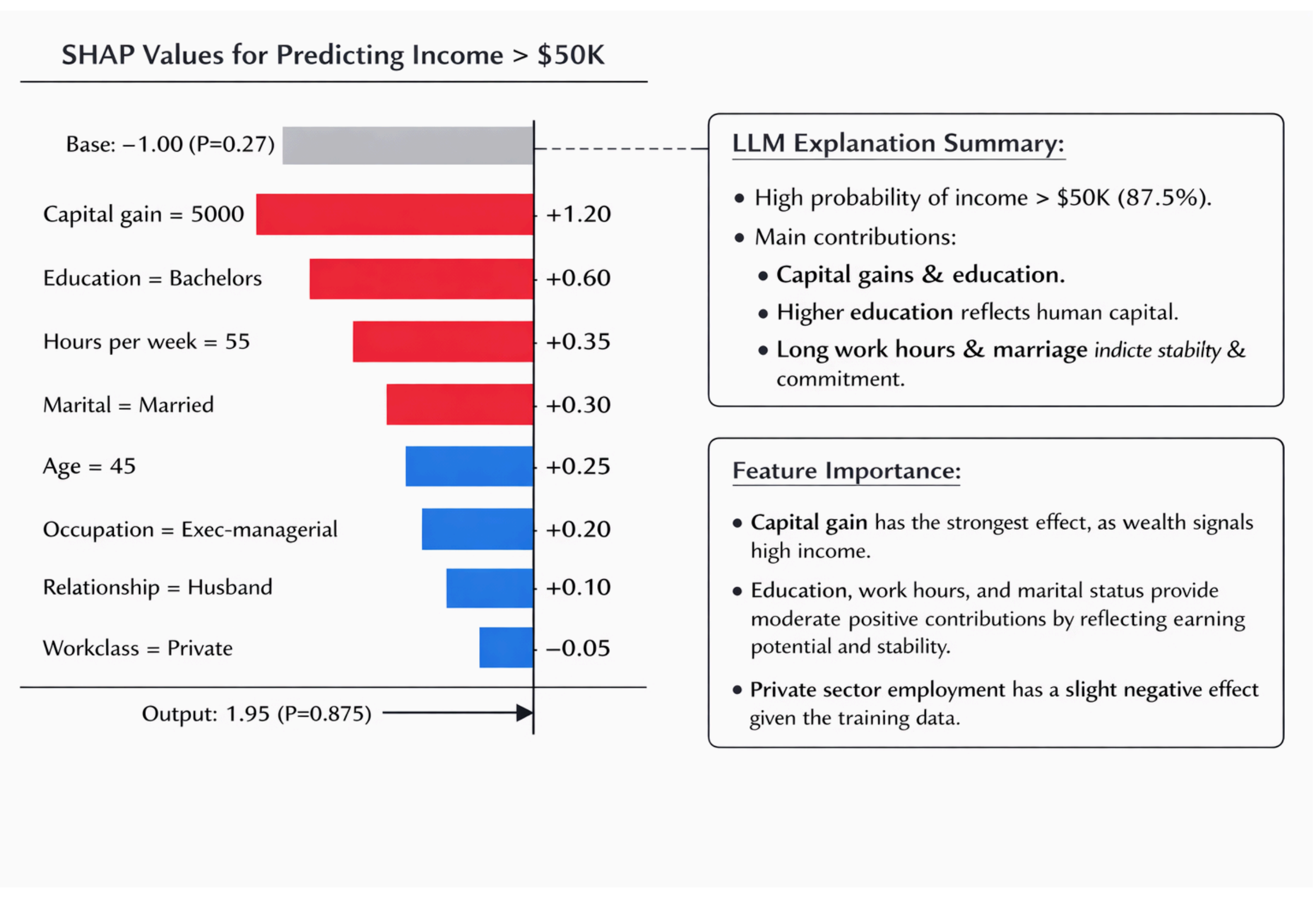

Figure 2. Example of Income > $50K Prediction Explained by SHAP Values and LLM Interpretation

## Discussion

The application of Large Language Models (LLMs) to enhance the interpretability of SHAP values marks a significant step toward making complex machine learning models more accessible to a broader audience. By translating technical outputs into plain language explanations, we address a critical gap between model developers and end users, particularly in high-stakes environments where understanding model decisions is paramount.

## Impact on Interpretability

Our approach demonstrates that LLMs, specifically the Mistral 7B model, can effectively bridge the interpretability gap by generating explanations that are both accurate and understandable. The use of the Titanic dataset as an example highlights how LLMs can simplify complex SHAP value outputs, transforming them into narratives that non-technical users can easily comprehend. This not only enhances trust in machine learning models but also empowers users to make informed decisions based on model predictions. The ability to present these explanations in natural language allows for a more intuitive understanding, which is crucial for users who may lack the technical expertise to interpret raw SHAP values.

## Limitations

While the results are promising, there are several limitations to our approach that warrant discussion. Firstly, the quality of the explanations generated by the LLM is heavily dependent on the quality of the prompts and the input data. Poorly structured prompts can lead to vague or misleading explanations, which could undermine the effectiveness of the approach. Additionally, LLMs like Mistral 7B are computationally intensive, requiring substantial resources for deployment and operation, particularly when processing large datasets or generating explanations in real-time. This could limit the practical applicability of our approach in resource-constrained environments.

Moreover, while our method enhances the accessibility of SHAP values, it does not entirely eliminate the need for technical oversight. There remains a risk that users might over-rely on these explanations without fully understanding the underlying model mechanics, potentially leading to misinterpretation of the model's decisions. It is also important to recognize that LLMs, including Mistral 7B, may occasionally produce incorrect or biased explanations, reflecting inherent biases in the training data or the model itself. These issues highlight the need for careful validation and possibly integration of additional safeguards to ensure the reliability of the explanations provided.

## Future Direction

Building on the foundation established in this paper, there are several promising avenues for future research and development that could further enhance the interpretability of SHAP values through LLMs:

- **Conduct a Small User Study**: A logical next step is to design a small-scale user study involving both technical and non-technical participants. This study would evaluate the clarity and usefulness of the LLM-generated explanations by gathering direct feedback from users. Such empirical evidence would provide valuable insights into the effectiveness of our approach and could form a strong contribution to the broader field of model interpretability.
- **Implement a Feedback Mechanism**: Incorporating a feedback loop where users can rate or comment on the explanations provided by the LLM offers a dynamic way to refine and improve the quality of the explanations over time. By adjusting the LLM's outputs based on user feedback, the system can become more adaptive to user needs, leading to more personalized and relevant explanations.

- **Fine-tuning the LLM on Domain-Specific Data**: To increase the relevance and accuracy of the explanations in specific contexts, we propose fine-tuning the LLM on domain-specific datasets, such as those in healthcare or finance. This would involve curating a dataset that includes both SHAP explanations and corresponding human-readable interpretations, allowing the LLM to generate more context-aware and precise explanations.
- **Propose a New Metric for Usability**: Another significant contribution could be the development of a new metric to measure the usability of LLM-generated explanations. This metric could be based on factors such as user comprehension, satisfaction, or decision accuracy. Establishing such a metric would provide a standardized way to evaluate and compare different interpretability methods, setting a benchmark for future research in this area.

These future directions not only aim to enhance the practical application of LLMs in improving SHAP interpretability but also seek to contribute novel insights and tools to the broader field of AI interpretability. By pursuing these avenues, we can continue to refine and expand the utility of LLMs in making complex machine learning models more accessible and trustworthy.

# Conclusion

In this paper, we have introduced the use of Large Language Models (LLMs) to enhance the interpretability of SHAP values, making complex machine learning models more accessible to non-technical users. By applying the Mistral 7B model, we successfully translated technical SHAP outputs into plain language explanations, thereby bridging the gap between model developers and end users. While our approach shows promise in improving clarity and trust in model predictions, it also highlights the importance of careful prompt engineering and the need for computational resources. Future work will focus on refining this approach through user studies, feedback mechanisms, and domain-specific fine-tuning, with the ultimate goal of advancing AI interpretability and accessibility.